\documentclass[twocolumn]{aastex631}
\usepackage{amssymb, amsmath}
\usepackage{graphicx}
\usepackage{natbib}
\usepackage{latexsym}
\usepackage{mathrsfs}
\usepackage{mathtools}
\usepackage{multirow}
\usepackage{xcolor}
\usepackage{color, colortbl}
\usepackage{booktabs,makecell}
\usepackage{float}
\usepackage{textcase}
\usepackage{censor}
\usepackage{array}
\usepackage{footmisc}

\newcommand{\tsup}{\textsuperscript}
\newcommand{\tsub}{\textsubscript}
\newcommand{\LTT}{LTT 1445A}
\newcommand{\Ha}{H$\alpha$}
\newcommand{\RpRs}{$R_\mathrm{p}/R_\mathrm{s}$}
\newcommand{\aRs}{$a/R_\mathrm{s}$}

\begin{document}

\title{Ground-based Optical Transmission Spectroscopy of the Nearby Terrestrial Exoplanet LTT 1445Ab}

\correspondingauthor{Hannah Diamond-Lowe}
\email{hdiamondlowe@space.dtu.dk}

\author[0000-0001-8274-6639]{Hannah Diamond-Lowe}
\affiliation{National Space Institute, Technical University of Denmark, Elektrovej 328, 2800 Kgs.\ Lyngby, Denmark}
\affiliation{Center for Astrophysics $\vert$ Harvard \& Smithsonian, 60 Garden St., Cambridge, MA 02138, USA}

\author[0000-0002-6907-4476]{Jo\~{a}o M.\ Mendon\c{c}a}
\affiliation{National Space Institute, Technical University of Denmark, Elektrovej 328, 2800 Kgs.\ Lyngby, Denmark}

\author[0000-0002-9003-484X]{David Charbonneau}
\affiliation{Center for Astrophysics $\vert$ Harvard \& Smithsonian, 60 Garden St., Cambridge, MA 02138, USA}

\author[0000-0003-1605-5666]{Lars A.\ Buchhave}
\affiliation{National Space Institute, Technical University of Denmark, Elektrovej 328, 2800 Kgs.\ Lyngby, Denmark}

\begin{abstract}
Nearby M dwarf systems currently offer the most favorable opportunities for spectroscopic investigations of terrestrial exoplanet atmospheres. The LTT~1445 system is a hierarchical triple of M dwarfs with two known planets orbiting the primary star, LTT~1445A. We observe four transits of the terrestrial world LTT~1445Ab ($R=1.3$ R$_\oplus$, $M=2.9$ M$_\oplus$) at low resolution with Magellan II/LDSS3C. We use the combined flux of the LTT~1445BC pair as a comparison star, marking the first time that an M dwarf is used to remove telluric variability from time-series observations of another M dwarf. We find H$\alpha$ in emission from both LTT~1445B and C, as well as a flare in one of the data sets from LTT~1445C. These contaminated data are removed from the analysis. We construct a broadband transit light curve of LTT~1445Ab from 620--1020 nm. Binned to 3-minute time bins we achieve an rms of 49 ppm for the combined broadband light curve. We construct a transmission spectrum with 20 spectrophotometric bins each spanning 20 nm and compare it to models of clear, 1$\times$ solar composition atmospheres. We rule out this atmospheric case with a surface pressure of 10 bars to $3.2 \sigma$ confidence, and with a surface pressure of 1 bar to $3.1 \sigma$ confidence. Upcoming secondary eclipse observations of LTT~1445Ab with JWST will further probe the cases of a high mean molecular weight atmosphere, a hazy or cloudy atmosphere, or no atmosphere at all on this terrestrial world.

\end{abstract}

\section{Introduction} \label{sec:intro}

The now decommissioned Kepler Space Telescope \citep{Borucki1997} stared at a single patch of sky for four years and taught us that worlds smaller than about 1.7 R$_\mathrm{\oplus}$ are common in the galaxy \citep{Fressin2013,Dressing2013,Dressing2015b}. But, the vast majority of the small worlds detected by Kepler are too faint for follow-up radial velocity measurements to determine a mass. The space-based Transiting Exoplanet Survey Satellite \citep[TESS;][]{Ricker2015}, along with ground-based surveys like MEarth \citep{Nutzman2008,Irwin2015} and TRAPPIST \citep{Gillon2013}, search the sky for small planets transiting the nearest and brightest stars. Four years into the TESS Mission the number of nearby worlds with $R<1.7\ \mathrm{R_\oplus}$ has grown into the tens, with many having mass measurements or upper limits. Based on both theoretical and empirical evidence \citep{Owen&Wu2013,Lopez2013,Rogers2015,Dressing2015a,Fulton2017,VanEylen2018,Luque&Palle2022}, we can definitively say that there exists a class of exoplanets with measured radii \textit{and} masses consistent with an Earth-like bulk composition. The question then becomes whether or not terrestrial exoplanets can possess substantial atmospheres and if so, what those atmospheres are comprised of.

Unambiguous evidence of an atmosphere around a terrestrial exoplanet has yet to be found. As such, determining whether not a terrestrial exoplanet has an atmosphere at all is a field of active development. Atmospheric circulation models predict that a thick atmosphere ($\gtrsim1$ bar of surface pressure) on a highly irradiated, tidally locked world will be able to advect energy from the sub-stellar point towards cooler longitudes \citep{Seager2009,Showman2013,Wordsworth2015,Koll2019}. The experiment is then to observe a phase curve or secondary eclipse of the terrestrial exoplanet and determine  if the planet's dayside thermal emission is reduced compared to the bare-rock case. 

Thick high surface pressure atmospheres efficiently redistribute heat from the dayside to the nightside; more tenuous, thin atmospheres do not. As such, low mean molecular weight hydrogen- and helium-dominated atmospheres with surface pressures below 10 bar are difficult to detect by their heat redistribution properties, but are accessible with transmission spectroscopy. Phase curve observations of the highly irradiated terrestrial world LHS 3844b resulted in a secondary eclipse depth consistent with a bare rock down to 10 bars of surface pressure \citep{Kreidberg2019}. Follow-up observations in transmission further ruled out low mean molecular weight atmospheres down to 0.1 bars \citep{Diamond-Lowe2020b}, providing further evidence that LHS 3844b is likely a bare rock.

We know from measuring the radii and masses of terrestrial exoplanets that extended hydrogen- and helium-dominated atmospheres making up $>1$\% the planet's mass are highly unlikely \citep{Owen2020b}, nor do terrestrial exoplanets possess a significant amount of water by mass that would shift their compositions away from what is measured for the Solar System terrestrial worlds \citep{Luque&Palle2022}. However, it is possible that small amounts ($<1\%$ of the planet's mass) of low mean molecular weight material can be retained from the protoplanetary nebula or out-gassed later on, and depending on the history of the terrestrial exoplanet, such an atmosphere may persist. Even a small amount of hydrogen and helium by mass is enough to increase a terrestrial exoplanet's scale height such that its atmosphere is detectable. 

For terrestrial worlds orbiting nearby ($<15$ pc) small ($<0.3$M$_\odot$) stars, low mean molecular weight atmospheres are accessible with low resolution transmission spectroscopy, when the planet passes in front of its host star from our line of sight. Employing this technique from both ground- and space-based observatories has delivered a growing number of results that find low mean molecular weight atmospheres on highly irradiated terrestrial worlds unlikely \citep{deWit2016,deWit2018,Diamond-Lowe2018,Diamond-Lowe2020b}, while still allowing for heavier, high mean molecular weight atmospheres or no atmospheres at all. 

In this work we focus on the terrestrial world \LTT b \citep{Winters2019}. \LTT b ($M_\mathrm{p} = 2.87^{+0.26}_{-0.25}\ \mathrm{M_\oplus}$; $R_\mathrm{p} = 1.305^{+0.066}_{-0.061}\ \mathrm{R_\oplus}$) orbits the primary star of a hierarchical triple M dwarf system with a period of $5.3587657^{+0.0000043}_{-0.0000042}$ days \citep{Winters2022}. The measured mass and radius place this world squarely on an Earth-like composition curve. At the time of writing, the \LTT\ system holds the distinction as being the closest M dwarf to host transiting planets, and given the remaining search space for such worlds, \LTT b is likely to remain one of the most spectroscopically accessible terrestrial worlds for transmission spectroscopy that we ever find. In thi  s work we present results from four transit observations captured with the Magellan II (Clay) telescope at the Las Campanas Observatory and the LDSS3C multi-object spectrograph. Our optical spectra range from 620--1020 nm and we take advantage of the rare case of being able to use the LTT 1445BC M dwarf binary as a comparison star with which to remove telluric variability during observations.

In Section~\ref{sec:obs} we explain our observing strategy. Section~\ref{sec:analysis} is dedicated to extracting and analyzing the spectroscopic data. In Section~\ref{sec:results} we present the main results of this work and discuss the implications for \LTT b's atmosphere. We also peer more closely at the LTT 1445BC binary companion and find a broadband flare in one data set and persistent variability in the H$\alpha$ line across all data sets. We wrap up with our conclusions in Section~\ref{sec:conclusion}.

\section{Observations} \label{sec:obs}

The terrestrial exoplanet \LTT b orbits about its host mid-M dwarf star every 5.36 days for a duration of 1.38 hours \citep{Winters2019,Winters2022}, which offers about five opportunities per year to observe a transit event from the Las Campanas Observatory in Chile. We were awarded time for four transit observations on the Magellan II (Clay) telescope in the 2019B semester (PI H.\ Diamond-Lowe). Two of these were lost due to bad weather. We were awarded a further four transits in the 2020B semester (PI D.\ Charbonneau), two of which were lost due to observatory closures related to the Covid-19 pandemic. From the 2019B and 2020B semesters we observed a total of four transits of \LTT b (Table~\ref{tab:obs}). We took spectra before, during, and after the planet transit for a total of 4.75 hours of observing time per-transit. Because we did not use a full night and were able to share the telescope with other programs, the observation time included a buffer of 15 minutes (7.5 minutes at the beginning and end of each observation) to change instruments at the telescope. 

To capture spectra of \LTT\ during the planet transit we used the multi-object Low Dispersion Survey Spectrograph (LDSS3C\footnote{\href{http://www.lco.cl/?epkb_post_type_1=ldss3_c-2}{LDSS3C Technical Specifications}}). We used an instrument set-up similar to previous programs that involved observing terrestrial exoplanets around nearby M dwarfs \citep{Diamond-Lowe2018,Diamond-Lowe2020a,Diamond-Lowe2020b}, however with key differences due to the brightness of \LTT\ \citep[V=11.2, T=8.88;][]{Henry2006,Winters2019} and to the fact that \LTT\ is part of a hierarchical triple system with two other mid-M dwarfs, LTT 1445B and C. 

\begin{figure}[!ht]
\includegraphics[width=0.47\textwidth]{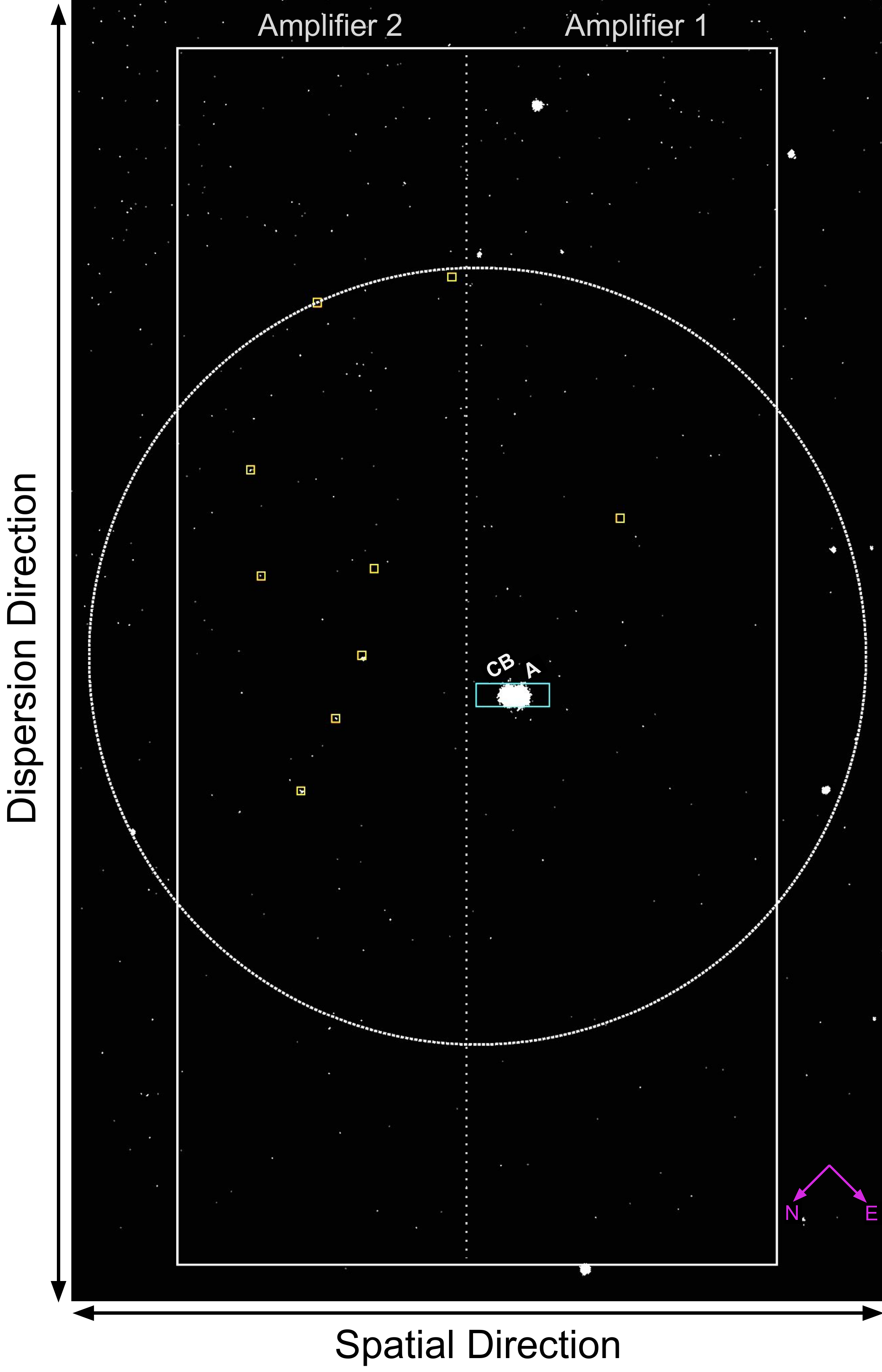}
\caption[]{Image of the LTT 1445 field in September 2019 from MEarth South \citep{Nutzman2008,Irwin2015}, which is available through the ExoFOP-TESS portal\footnote{\href{https://exofop.ipac.caltech.edu/tess/index.php}{exofop.ipac.caltech.edu/tess/index.php}}. The large circle shows the field of view of LDSS3C, and the solid rectangle is the extent of the LDSS3C detector. The two LDSS3C amplifiers are labeled at the top of the detector rectangle, with the vertical dotted line indicating the split on the detector. The light blue rectangle is the $47''\times15''$ slit for LTT 1445ABC. Orange squares indicate small slits for alignment stars.}
\label{fig:onsky}
\end{figure}

A multi-object spectrograph where the slit sizes are adjustable is required for low resolution ground-based time-series observations like these because comparison stars must be observed simultaneously with a target star in order to remove telluric variations imprinted on the stellar spectra during the observations. Comparison stars should be of comparable magnitude to the target star in the bandpass of observations, which means that usually M dwarfs are compared to spectrally different FGK stars. In this case, the combined spectra of the companion M dwarfs LTT 1445BC can be used as a comparison star to \LTT, marking the first time that an M dwarf can be compared to another M dwarf in multi-object spectroscopy for the purposes of constraining a planetary atmosphere. We note that given the alignment of the LTT 1445 stars we are effectively using a single slit as opposed to multiple slits (Figure~\ref{fig:onsky}).

We place all three stars in the LTT 1445 system in the same slit on the science mask, $47''$ wide in the cross-dispersion direction and $15''$ in the dispersion direction (Figure~\ref{fig:onsky}). This allows us to measure $20''$ of sky background on either side of \LTT\ and the LTT1445BC pair, as well as enough room in the dispersion direction to guard against light losses. The calibration mask is identical to the science mask but with $0.5''$ slits in the dispersion direction. During the afternoon prior to observations we take biases, darks, and quartz flats with the science mask, as well as helium, neon, and argon arcs with the calibration mask. During night-time observations we take a set of undispersed reference images with the science mask before and after the main science images in order to mark the location of the LTT 1445 stars on the detector. These undispersed images with the science mask are useful during data extraction to mark the location of the stars on the detector.

With LDSS3C we use the \texttt{1x1} detector binning, \texttt{Turbo} readout speed, and \texttt{Low} gain settings. These settings accommodate the brightness of the LTT 1445 stars. According to the LDSS3C User Manual\footnote{\href{http://www.lco.cl/?epkb_post_type_1=ldss-3-user-manual}{LDSS3C User Manual}}, the \texttt{Turbo} readout speed is not supported for science observations due to variable readout noise that can adversely affect data taken for faint sources. For this mode the read noise is 8 e$^-$ rms \citep{Stevenson2016a}. In the case of the LTT 1445 stars we are far from the read noise limit, so we use the \texttt{Turbo} readout mode to improve the observing cadence \citep[][A.\ Seifahrt, priv.\ comm.]{Stevenson2016a}. For most of the observations we took 15 s exposures with 28.5 s readout time, bringing the duty cycle to 34.5\%. At the start of observations for Data Set 3 seeing conditions were poor so we increased the exposure time to 20 s, however during observations seeing improved dramatically and we reduced the exposure time to 15 s shortly after transit egress in order to avoid saturating any detector pixels (Table~\ref{tab:obs}). 

LDSS3C's 16-bit analog-to-digital converter (ADC) has a saturation limit of 65,535 analog-to-digital units (ADUs). We ensure that all pixels used in the data analysis stay under this limit during observations. Given that the gains listed in the LDSS3C User Manual are out of date, the LDSS3C instrument specialists at Las Campanas Observatory made updated gain measurements in each of the two CCD amplifiers during the observing season. For the 2019 observations we use gains of 2.95 e$^-$/ADU and 2.56 e$^-$/ADU for amplifiers 1 and 2, respectively, to convert the measured ADUs to photoelectrons. For the 2020 observations we use gains of 2.84 e$^-$/ADU and 2.52 e$^-$/ADU, respectively.

\begin{deluxetable*}{ccccccccc}
\centering
\caption{Observations with Magellan II (Clay) and the LDSS3C Multi-object Spectrograph\label{tab:obs}}
\tablewidth{0pt}
\tablehead{
\colhead{Data Set} & \colhead{Transit} & \colhead{Night} & \colhead{Time} & \colhead{Exp.\ Time} & \colhead{Duty Cycle} & \colhead{Number of} & \colhead{Airmass} & \colhead{Seeing\tsup{b}}\\
\colhead{Number}   & \colhead{Number\tsup{a}} & \colhead{ (UTC) }    & \colhead{(UTC)} & \colhead{(s)}           & \colhead{(\%) }& \colhead{Exposures} & \colhead{at $t_0$}        & \colhead{(arcsec)}
}
\startdata
\tsup{c} & 56  & 2019-08-17 & --------------------- & --- & --- & --- & ------ & --- \\
\tsup{c} & 67  & 2019-10-15 & --------------------- & --- & --- & --- & ------ & --- \\
1        & 75  & 2019-11-27 & 01:08:29--05:04:36  & 15  & 34.5 & 326 & 1.192 & 0.70--1.10 \\
2        & 78  & 2019-12-13 & 02:34:15--06:46:19  & 15  & 34.5 & 350 & 1.025 & 0.60--0.80 \\
\tsup{d} & 131 & 2020-09-22 & --------------------- & --- & --- & --- & ------ & --- \\
\tsup{d} & 134 & 2020-10-08 & --------------------- & --- & --- & --- & ------ & ---\\
3        & 142 & 2020-11-20 & 01:40:32--06:05:24 & 20, 15 & 41.1, 34.5 & 215, 127 & 1.176 & 1.00--0.45 \\
4        & 145 & 2020-12-06 & 03:28:42--07:39:04 & 15  & 34.5 & 343 & 1.042 & 0.50--0.60 \\
\enddata
\tablecomments{\\
\tsup{a} Transit number is counted from the transit ephemeris $T_0 = 2458412.70851$ \citep{Winters2022}. \\
\tsup{b} Seeing recorded at the beginning and end of observations. Conditions varied during observations 1, 2, and 3.\\
\tsup{c} No data due to bad weather.\\
\tsup{d} No data due to telescope closure in response to the Covid-19 pandemic.\\
}
\end{deluxetable*}

\section{Data Extraction \& Analysis}\label{sec:analysis}

\begin{figure}
\includegraphics[width=0.48\textwidth]{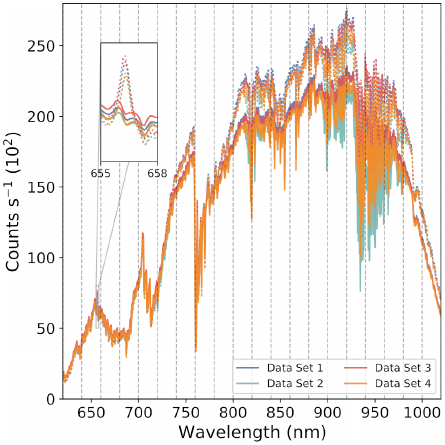}
\caption{Average spectra of \LTT\ (solid lines) and LTT 1445BC combined (dotted lines) from each of the 4 data sets. The spectra represent the total counts from each star measured over the course of each observing night, normalized by the total integration time (exposure time $\times$ number of exposures). Vertical gray lines are the edges of the 20 nm spectrophotometric bands. The inset zooms in on the \Ha\ line, which is evident in emission in the combined LTT 1445BC comparison star, but not in \LTT.}
\label{fig:spectrabinned}
\end{figure}

Our data extraction and analysis steps are nearly identical to those outlined in \citet{Diamond-Lowe2020a, Diamond-Lowe2020b}. We use the \texttt{mosasaurus}\footnote{\href{https://github.com/zkbt/mosasaurus}{github.com/zkbt/mosasaurus}} extraction pipeline to turn the raw images taken at the telescope into data cubes that include time series of wavelength calibrated 1D spectra for each star in each data set (we treat LTT 1445BC as a single star in the extraction; Figure~\ref{fig:spectrabinned}). We also record auxiliary information collected during the observations and subsequent extraction, for example the airmass and instrument rotation angle at the time of each integration and the average widths of the stellar spectral traces on the detector over the course of the time series, which is a proxy for seeing conditions during observations. 

Using the LTT 1445BC pair as the comparison star yields two main differences as compared to previous extraction and analysis steps described in \citet{Diamond-Lowe2020a,Diamond-Lowe2020b}. The first occurs during wavelength calibration. After an initial wavelength solution is found using calibration arcs we then perform a finer calibration where we cross-correlate elements of each spectrum with a master spectrum. This resulting wavelength correction is determined for every spectrum and for every data set. Usually the cross-correlation step would use telluric features (e.g., oxygen and water lines) as well as prominent stellar features (e.g., Ca \textsc{II} triplet lines) to finely align the spectra. In this case, LTT 1445BC is the same spectral type as the target star \LTT\ so we can cross-correlate against the entire spectrum, not just a subset of features. We divide the spectra up into 10 chunks in order to allow for both a shift and stretch in the spectra during the course of observations. Using the LTT 1445BC M dwarf pair as a comparison star provides a marked advantage during wavelength calibration, especially at the edges of the spectra where low counts can make cross-correlating with telluric features difficult.

The second difference in this data set is that the LTT 1445BC time series exhibits variability around the \Ha\ line, which is not the case for \LTT (Figure~\ref{fig:spectrabinned}). Because this variability is not telluric but rather inherent to one or both of the LTT 1445BC components, we cannot use any spectral information around the \Ha\ line when using LTT 1445BC to remove broad telluric variability from the \LTT\ time series. We therefore exclude data between 655--658 nm when performing the analyses. In Data Set 1 we also capture a broadband flare in the LTT 1445BC component. We exclude the four time points associated with this flare from the broadband and spectrophotometric analysis of Data Set 1. The flare and \Ha\ variability in the BC component is further discussed in Section~\ref{subsec:Ha_variability}.

We construct broadband light curves by summing all of the flux (with the exception of the \Ha\ line) from 620--1020 nm. We fit the light curve using a Gaussian process (GP) regression with a Mat\'ern $3/2$ kernel \citep[\texttt{george};][]{Ambikasaran2015} that takes a transit model \citep[\texttt{batman};][]{Kreidberg2015} as the mean function. For each data set we fit for the time of mid-transit $t_0$, planet-to-star radius ratio \RpRs, the cosine of the inclination cos $i$, the scaled semi-major axis \aRs, and two logarithmic limb-darkening parameters $l_0$ and $l_1$, which are re-parameterized according to \citet{Espinoza2016}. We place Gaussian priors on the re-parameterized limb-darkening coefficients set by the mean and five times the uncertainty given by the Limb Darkening Toolkit \citep[\texttt{LDTk};][]{Parviainen2015}. We assume a circular orbit ($e=0$) since stand-alone transits like the ones we present here provide negligible information on a planet's eccentricity.

\begin{figure}[!ht]
\includegraphics[width=0.47\textwidth]{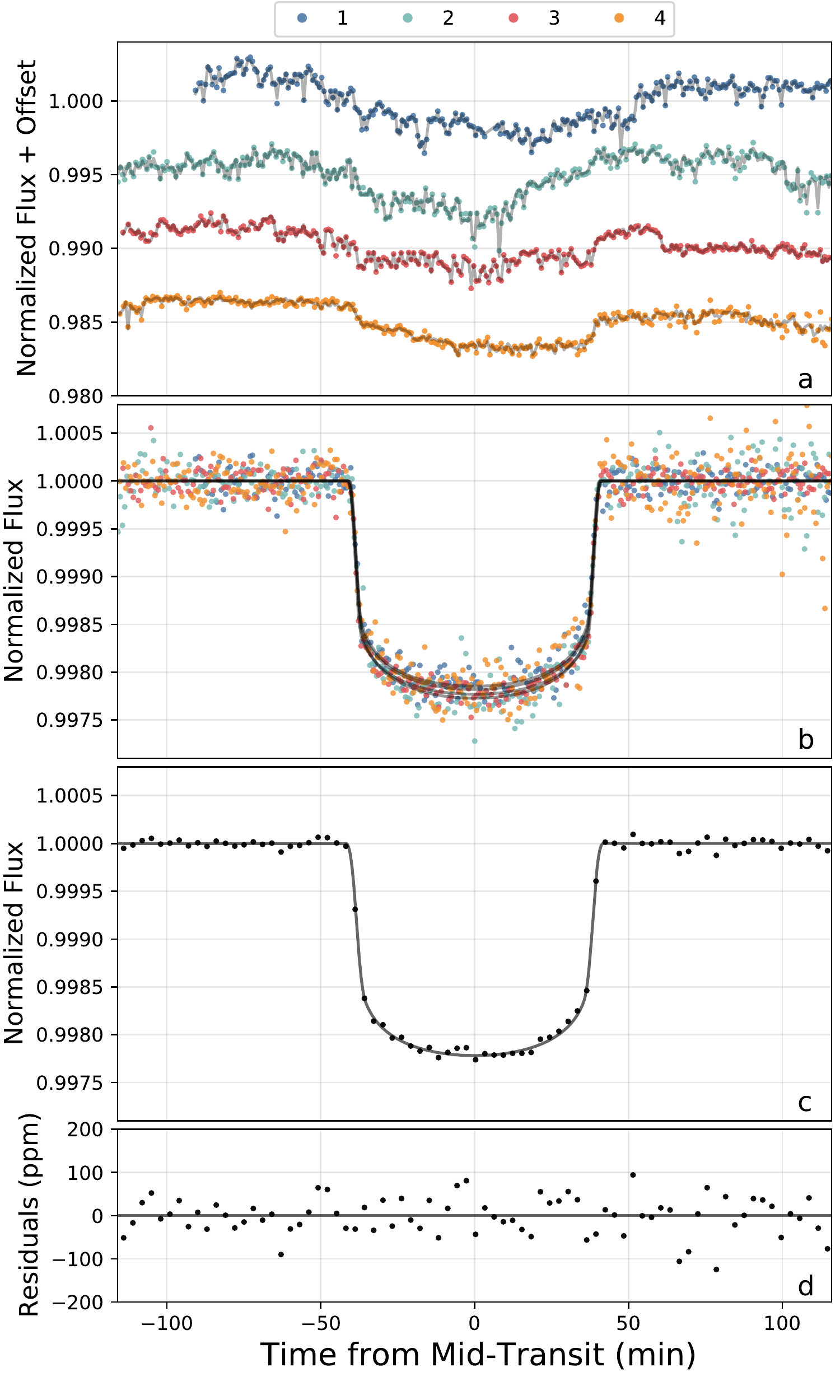}
\caption{Panel a: Broadband light curves of four transit observations of \LTT b. Each color represents a different data set, corresponding to the data set numbers in Table~\ref{tab:obs}. The combined transit and GP noise model is shown in grey. The raw light curves are highly correlated with the width and centroid vectors from the LTT 1445BC spectral trace. These regressors, preferred in every data set, are responsible for the strong variation in the GP noise models. Panel b: Same as panel a but with the GP noise component removed. The rms of the combined light curve is 200 ppm. Panel c: Same as panel b but with the combined light curve binned to 3-minute time bins. The transit model is smoothed using a 3-minute box-car kernel. The rms of the 3 min-binned combined light curve is 49 ppm. Panel d: Residuals of panel c.}
\label{fig:broadlc}
\end{figure}

We use a dynamic multi-nested sampler \citep[\texttt{dynesty};][]{Speagle2020} to explore the parameter space. Following \citet{Diamond-Lowe2020b}, we test several vectors containing auxiliary information, such as airmass, centroid, etc., to use as regressors in the GP, and select the best combination according to the Bayesian evidence output from \texttt{dynesty}. The regressors used for each data set can be found in Table~\ref{tab:regressors}. The best-fit values and figures from the broadband light curve analysis can be found in Table~\ref{tab:broadlc} and Figure~\ref{fig:broadlc}. Our best fit orbital parameters from the broadband fit are in excellent agreement with those of \citet{Winters2022}. Since the initial preparation of this manuscript, \citet{Lavie2022} provided an update to the orbital parameters of \LTT b using both TESS transit data and VLT/ESPRESSO radial velocity data. There is some discrepancy between their reported orbital parameters for \LTT b and those of \citet{Winters2022} and this work. Further analysis beyond the scope of this work is necessary to resolve these discrepancies.

The spectra of LTT 1445BC are not completely resolved on the detector so we use the combined spectrum of this binary pair as a comparison star to \LTT. We note that the resulting light curves are highly correlated with the changing width and centroid of the LTT 1445BC spectral trace, likely because \texttt{mosasaurus} is built to fit one Gaussian profile instead of two\footnote{In attempting to separate the B and C components we fitted a double Gaussian profile to the LTT 1445BC trace, but were unable to achieve satisfactory results.}. When fitting the light curves with a Gaussian process the width and centroid vectors are unsurprisingly favored as regressors in every data set. We also do not necessarily need to include airmass as GP regressor since there is negligible differential extinction between the comparison and target stars (though Data Sets 2 and 4 still prefer the use of airmass as a regressor).

\begin{figure*}
\includegraphics[width=0.96\textwidth]{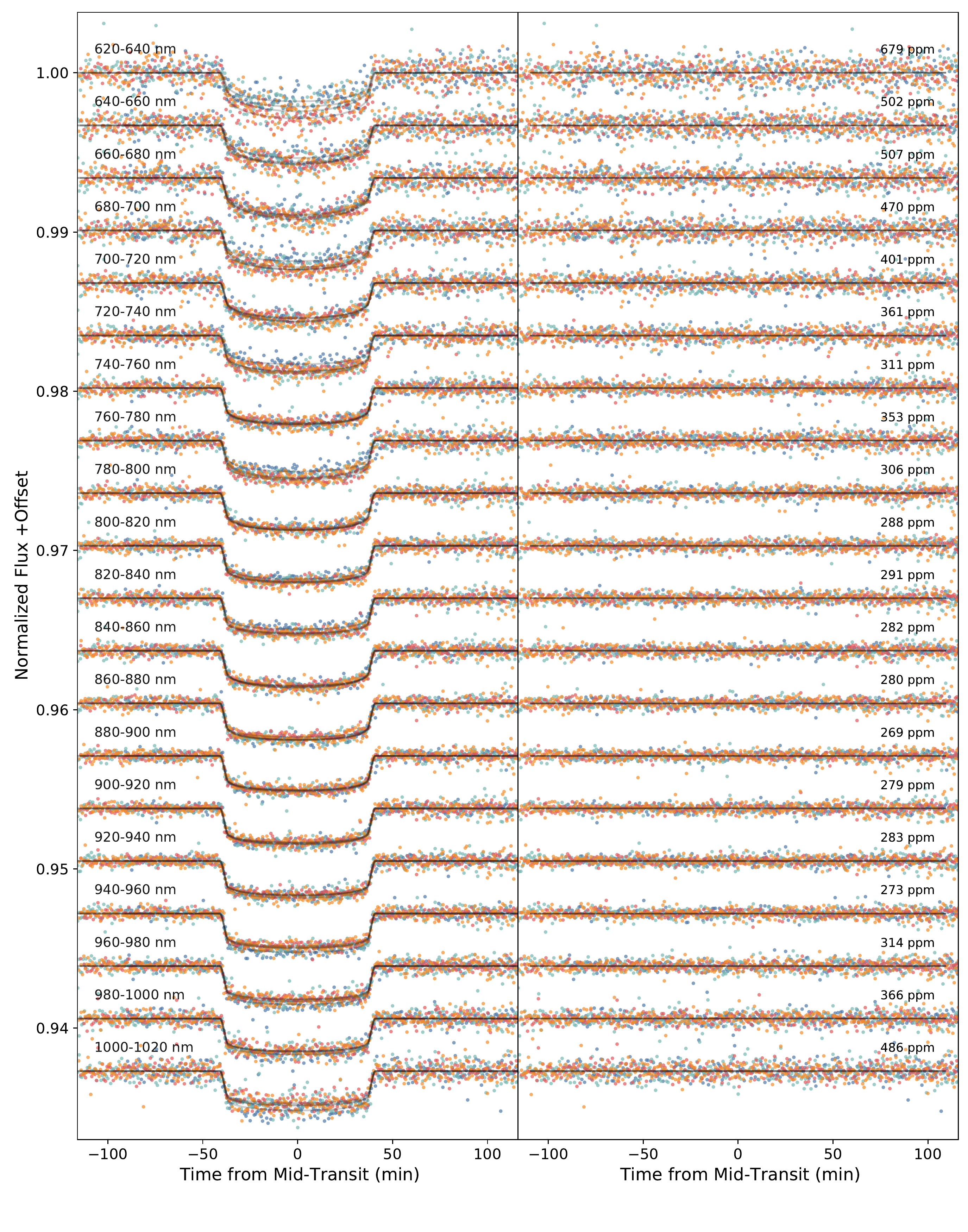}
\vspace{-25pt}
\caption{Left: spectroscopic light curves with the GP noise component removed for each of the 4 data sets. The transit model (GP mean function) is plotted in gray for each data set. Light curves are offset for clarity. Right: residuals compared to the best fit transit model for each of the 4 data sets along with the combined rms. The individual transit depths and uncertainties are provided in Table~\ref{tab:transitdepths} and shown in Figure~\ref{fig:indvspeclcs}.}
\label{fig:speclcs}
\end{figure*}

For each data set we divide the broadband spectra into 20 nm spectrophotometric bands and sum up the flux in each band to create 20 spectral light curves. We perform a GP regression on each of these spectral light curves, allowing the hyperparameters to vary but using the same set of regressors as for the broadband light curves. For the spectral light curves we fix $t_0$, cos $i$, and \aRs\ to the best value from the broadband light curve fit and then fit for \RpRs, $l_0$, and $l_1$ for each data set. We provide the resulting spectrophotometric transit depths for each data set, as well as the inverse-variance weighted average transit depths, in Table~\ref{tab:transitdepths} and Figure~\ref{fig:speclcs}.

\definecolor{Gray}{gray}{0.85}
\newcolumntype{g}{>{\columncolor{Gray}[1.5mm][1.5mm]}c}
%\begin{tabular}
\begin{deluxetable}{lgcgc}
\tablecaption{GP Regression Data\label{tab:regressors}}
\tablewidth{0pt}
%\hline
%\hline
\tablehead{
\colhead{Input} &\multicolumn{4}{c}{Data Set Number}\\
\cline{2-5}
\colhead{Regressors} & \colhead{1} & \colhead{2} & \colhead{3} & \colhead{4}
}
\startdata
Airmass        &            & \checkmark &            & \checkmark \\
Rotation angle & \checkmark & \checkmark &            & \checkmark \\
Centroid       & \checkmark & \checkmark & \checkmark & \checkmark \\
Width          & \checkmark & \checkmark & \checkmark & \checkmark \\
Peak           &            &            & \checkmark &            \\
Shift          &            &            & \checkmark & \checkmark \\
Stretch        &            &            &            &            \\
Time           &            &            & \checkmark &            \\
\enddata
\tablecomments{A more detailed explanation of these regressors can be found in \citet{Diamond-Lowe2020a}.}
\end{deluxetable}

\begin{deluxetable*}{lc|cccc}
\tablecaption{Broadband light curve transit parameters and priors\label{tab:broadlc}}
\tablewidth{0pt}
\tablehead{
\multirow{2}{*}{Parameter} & \multirow{2}{*}{Prior} &\multicolumn{4}{c}{Data Set Number} \\
& & \colhead{1} & \colhead{2} & \colhead{3} & \colhead{4}
}
\startdata
$\delta t_0$ (days) $^a$ & $\mathcal{U}$(-0.001, 0.001) $^b$  & -0.00019 $\pm$ 0.00009 & -0.00025 $\pm$ 0.00011 & 0.00015 $\pm$ 0.00007 & 0.00021 $\pm$ 0.00016 \\ 
$R_p/R_s$ & $\mathcal{U}(0.02, 0.07)$ $^b$ & 0.0432 $\pm$ 0.0010 & 0.0447 $\pm$ 0.0009 & 0.0443 $\pm$ 0.0008 & 0.0439 $\pm$ 0.0008 \\ 
cos\ $i$ & $\mathcal{U}(0.0, 1.0)$ $^b$ & $0.0052^{+0.0045}_{-0.0031}$ & $0.0094^{+0.0056}_{-0.0047}$ & $0.0059^{+0.0046}_{-0.0036}$ & $0.0118^{+0.0079}_{-0.0068}$ \\ 
$a/R_\mathrm{s}$ & $\mathcal{U}(5, 100)$ $^b$ & $31.33^{+0.37}_{-0.95}$ & $30.39^{+0.99}_{-1.76}$ & $31.00^{+0.45}_{-1.05}$ & $29.62^{+1.59}_{-2.81}$ \\ 
$l_0$ & $\mathcal{N}(0.43, 0.19)$ $^b$ & 0.42 $\pm$ 0.17 & 0.44 $\pm$ 0.15 & 0.53 $\pm$ 0.15 & 0.46 $\pm$ 0.16 \\ 
$l_1$ & $\mathcal{N}(0.58, 0.06)$ $^b$ & 0.568 $\pm$ 0.052 & 0.573 $\pm$ 0.044 & 0.619 $\pm$ 0.042 & 0.572 $\pm$ 0.047 \\ 
 \multicolumn{2}{c}{RMS (ppm) $^c$}   & 124 & 207 & 115 & 289 \\ 
\enddata
\tablecomments{\\
\tsup{a} $\delta t_0$ is the difference between the predicted time of mid-transit and the derived time of mid-transit from fitting each broadband light curve. The predicted time of mid-transit is calculated as $t_0 = T_0 + nP$ where $T_0 = 2458412.70851\ \mathrm{BJD_{TDB}}$ and $P = 5.3587657$ days \citep{Winters2022}; $n$ is the transit number given in Table~\ref{tab:obs} for each transit. All resulting $\delta t_0$ values are less than the observation cadence time of approximately 48 s per integration (15 or 20 s exposure time plus readout time).\\
\tsup{b} $\mathcal{U}$ stands for uniform prior; $\mathcal{N}$ stands for Gaussian prior.\\
\tsup{c} The rms values are calculated by comparing the unbinned broadband light curve residuals to the GP model for each data set.}
\end{deluxetable*}

\definecolor{Gray}{gray}{0.85}
\begin{deluxetable*}{c|ccccggg}
\tablecaption{Spectroscopic Transit Depths\label{tab:transitdepths}}
\tablewidth{0pt}
\tablehead{
\rowcolor{white}\colhead{Wavelength} & \multicolumn{4}{c}{Transit Depths $(R_\mathrm{p}/R_\mathrm{s})^2$ by Data Set (\%)} & \colhead{Mean $(R_\mathrm{p}/R_\mathrm{s})^2$} & \multirow{2}{*}{rms} & \colhead{$\times$ Exp.} \\
\rowcolor{white} \colhead{(nm)}& \colhead{1} & \colhead{2} & \colhead{3} & \colhead{4} & \colhead{(\%)} & \colhead{(ppm)} & \colhead{Noise}
}
\startdata
620-640 & 0.1434 $\pm$ 0.0393 & 0.1734 $\pm$ 0.0192 & 0.2340 $\pm$ 0.0443 & 0.1875 $\pm$ 0.0117 & 0.1836 $\pm$ 0.0095 & 680 & 1.32 \\ 
640-660\tsup{a} & 0.1623 $\pm$ 0.0198 & 0.1920 $\pm$ 0.0125 & 0.2003 $\pm$ 0.0233 & 0.1883 $\pm$ 0.0094 & 0.1874 $\pm$ 0.0067 & 502 & 1.31 \\ 
660-680 & 0.1826 $\pm$ 0.0263 & 0.2041 $\pm$ 0.0139 & 0.1951 $\pm$ 0.0339 & 0.1994 $\pm$ 0.0099 & 0.1992 $\pm$ 0.0075 & 508 & 1.37 \\ 
680-700 & 0.1599 $\pm$ 0.0227 & 0.1874 $\pm$ 0.0109 & 0.2071 $\pm$ 0.0251 & 0.2000 $\pm$ 0.0090 & 0.1929 $\pm$ 0.0064 & 471 & 1.35 \\ 
700-720 & 0.1860 $\pm$ 0.0184 & 0.1949 $\pm$ 0.0098 & 0.1829 $\pm$ 0.0207 & 0.2001 $\pm$ 0.0085 & 0.1955 $\pm$ 0.0058 & 402 & 1.38 \\ 
720-740 & 0.1465 $\pm$ 0.0180 & 0.2028 $\pm$ 0.0077 & 0.1922 $\pm$ 0.0131 & 0.1904 $\pm$ 0.0073 & 0.1923 $\pm$ 0.0047 & 365 & 1.53 \\ 
740-760 & 0.1949 $\pm$ 0.0112 & 0.1966 $\pm$ 0.0074 & 0.1894 $\pm$ 0.0065 & 0.2002 $\pm$ 0.0069 & 0.1950 $\pm$ 0.0037 & 313 & 1.55 \\ 
760-780 & 0.1608 $\pm$ 0.0161 & 0.2016 $\pm$ 0.0082 & 0.2130 $\pm$ 0.0136 & 0.1930 $\pm$ 0.0079 & 0.1958 $\pm$ 0.0050 & 355 & 1.51 \\ 
780-800 & 0.1893 $\pm$ 0.0117 & 0.1974 $\pm$ 0.0074 & 0.1999 $\pm$ 0.0088 & 0.1993 $\pm$ 0.0073 & 0.1976 $\pm$ 0.0042 & 308 & 1.52 \\ 
800-820 & 0.1850 $\pm$ 0.0091 & 0.1966 $\pm$ 0.0064 & 0.2026 $\pm$ 0.0064 & 0.2050 $\pm$ 0.0066 & 0.1989 $\pm$ 0.0034 & 290 & 1.55 \\ 
820-840 & 0.1703 $\pm$ 0.0106 & 0.2024 $\pm$ 0.0065 & 0.1964 $\pm$ 0.0071 & 0.1936 $\pm$ 0.0065 & 0.1943 $\pm$ 0.0036 & 297 & 1.63 \\ 
840-860 & 0.1892 $\pm$ 0.0100 & 0.1963 $\pm$ 0.0058 & 0.1937 $\pm$ 0.0128 & 0.1983 $\pm$ 0.0075 & 0.1955 $\pm$ 0.0040 & 283 & 1.55 \\ 
860-880 & 0.2009 $\pm$ 0.0088 & 0.2038 $\pm$ 0.0064 & 0.1904 $\pm$ 0.0074 & 0.1998 $\pm$ 0.0068 & 0.1990 $\pm$ 0.0036 & 282 & 1.60 \\ 
880-900 & 0.2090 $\pm$ 0.0079 & 0.1941 $\pm$ 0.0059 & 0.1933 $\pm$ 0.0059 & 0.1880 $\pm$ 0.0061 & 0.1946 $\pm$ 0.0031 & 270 & 1.55 \\ 
900-920 & 0.2019 $\pm$ 0.0095 & 0.2013 $\pm$ 0.0059 & 0.1942 $\pm$ 0.0066 & 0.1906 $\pm$ 0.0061 & 0.1964 $\pm$ 0.0033 & 280 & 1.63 \\ 
920-940 & 0.1979 $\pm$ 0.0104 & 0.1950 $\pm$ 0.0065 & 0.1897 $\pm$ 0.0057 & 0.2056 $\pm$ 0.0063 & 0.1965 $\pm$ 0.0033 & 285 & 1.55 \\ 
940-960 & 0.2216 $\pm$ 0.0051 & 0.1967 $\pm$ 0.0059 & 0.1851 $\pm$ 0.0053 & 0.1912 $\pm$ 0.0062 & 0.1998 $\pm$ 0.0028 & 275 & 1.42 \\ 
960-980 & 0.2200 $\pm$ 0.0122 & 0.1904 $\pm$ 0.0098 & 0.1881 $\pm$ 0.0077 & 0.2018 $\pm$ 0.0079 & 0.1972 $\pm$ 0.0044 & 315 & 1.61 \\ 
980-1000 & 0.2045 $\pm$ 0.0166 & 0.1850 $\pm$ 0.0082 & 0.1830 $\pm$ 0.0083 & 0.1903 $\pm$ 0.0075 & 0.1877 $\pm$ 0.0044 & 366 & 1.73 \\ 
1000-1020 & 0.2221 $\pm$ 0.0072 & 0.1911 $\pm$ 0.0103 & 0.1803 $\pm$ 0.0107 & 0.1950 $\pm$ 0.0090 & 0.2022 $\pm$ 0.0045 & 486 & 1.80 \\ 
\enddata
\tablecomments{The final three columns in grey provide the inverse-variance weighted mean across all 4 data sets for each spectroscopic band, along with the RMS of the 4 data sets combined in each band. The final column, $\times$ Expected Noise, describes how close, on average, we get to the calculated photon noise in each band.\\
\tsup{a} This spectral band excludes wavelengths covering the \Ha\ line (655--658 nm)}
\end{deluxetable*}

\begin{figure*}
\includegraphics[width=\textwidth]{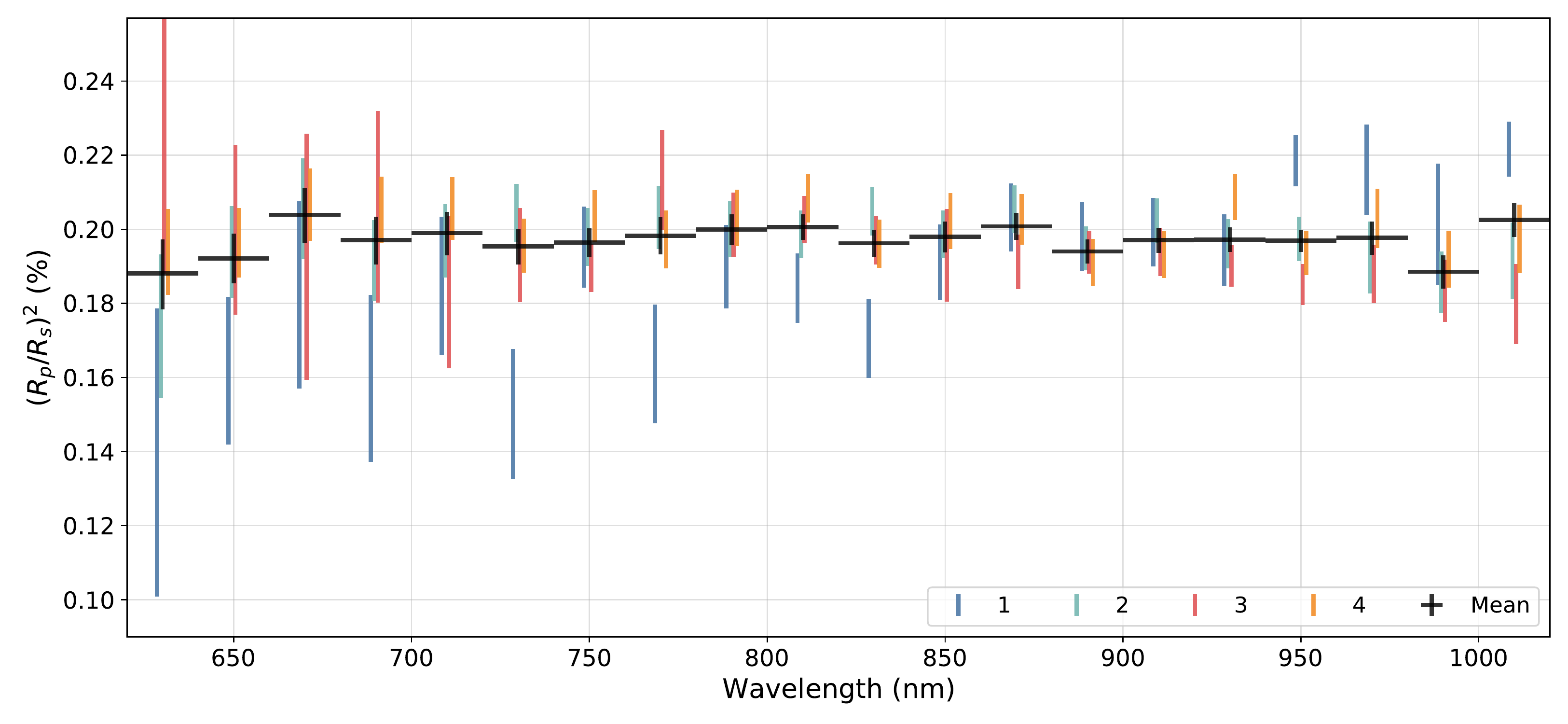}
\caption{Transmission spectra from Data Sets 1--4. The spectroscopic data are binned to 20 nm and the individual transmission spectra are offset for clarity in the x-axis. Bars in the y-axis represent 68\% confidence intervals, with black bars representing the inverse-variance weighted mean value of $(R_\mathrm{p}/R_\mathrm{s})^2$ in each bin; black bars along the x-axis illustrate the extent of each bin in wavelength.}
\label{fig:indvspeclcs}
\end{figure*}

\section{Results \& Discussion}\label{sec:results}

\subsection{Transmission spectrum of LTT 1445Ab compared to atmospheric models}

To create our transmission spectrum, we take the inverse-variance weighted mean of the transit depths in each spectrophotometric band. We can then compare this observed transmission spectrum to model transmission spectra of \LTT b to test the likelihood of a set of 1$\times$ solar composition ($\mu=2.3$ g mol\tsup{-1}) atmospheric cases with surface pressures ranging from 0.01--10.0 bars. 

To build the model transmission spectra we first determine temperature-pressure (T-P) profiles for each atmospheric case using a 1D radiative-convective model developed in \cite{Mendonca2020}. The code uses a 2-stream formulation with multiple scattering effects to represent the radiative processes in the atmosphere \citep{Mendonca2015}. To ensure that the vertical gradient of the temperature does not exceed the adiabatic profile, we use a simple convective adjustment scheme to mix the enthalpy instantaneously in a buoyant unstable atmospheric region \citep{Mendonca2018,Malik2019a}. Our radiative transfer code uses an opacity ($\kappa$) table with a correlated-$\kappa$ approximation as described in \citet{Malik2017} and the same parameters as in \citet{Malik2019b}. The cross-sections in the $\kappa$-distribution table are computed using the \texttt{HELIOS-K} software \citep{Grimm2021} from the following line lists: H$_2$O \citep{Barber2006}, CO$_2$ \citep{Rothman2010}, CO \citep{Li2015}), CH$_4$ \citep{Yurchenko2014}), NH$_3$ \citep{Yurchenko2011}, HCN \citep{Harris2006}, C$_2$H$_2$ \citep{Gordon2017}, PH$_3$ \citep{Sousa-Silva2015}, H$_2$S \citep{Azzam2016}, Na and K \citep{Burrows2000,Burrows2003,Kurucz2011}; collision-induced absorption (CIA) of H$_2$-H$_2$ \citep{Richard2012} and H$_2$-He \citep{Richard2012}; and scattering cross-sections of H$_2$ \citep{Sneep2005} and H \citep{Lee2004}. 

We calculate the chemical concentrations with the open-source code \texttt{FastChem} \citep{Stock2018}, with a constant surface albedo set to 0.2. All simulations are integrated until radiative-convective equilibrium is reached \citep{Malik2017}. We then feed these custom T-P profiles into the publicly available \texttt{Exo-Transmit} code \citep{Miller-RicciKempton2012,Kempton2017} in order to generate model transmission spectra. Similar to the use of model transmission spectra in \citet{Diamond-Lowe2018,Diamond-Lowe2020a,Diamond-Lowe2020b}, we allow the surface radius of \LTT b to vary in the model until we get a result that most closely matches the observed transmission spectrum based on the reduced $\chi^2$ statistic, thereby accounting for our uncertainty in the true radius at the solid planet surface.

\begin{figure*}
\includegraphics[width=\textwidth]{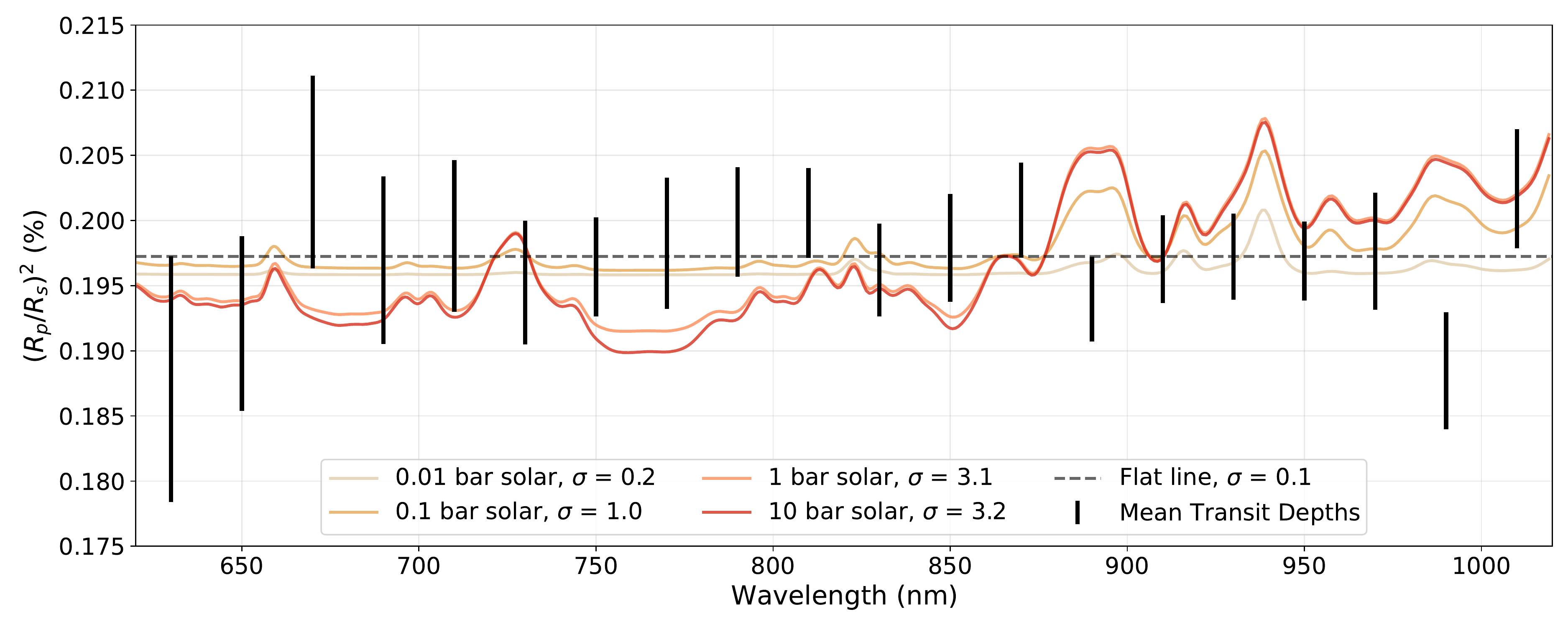}
\caption{Observed transmission spectrum of \LTT b from four data sets combined (black points the same as in Figure~\ref{fig:indvspeclcs}) compared to model transmission spectra derived from 1$\times$ solar composition atmospheres from 0.01--10 bars of surface pressure. The $\sigma$ values in the legend are the confidence to which we can rule out the corresponding model transmission spectrum. We confidently rule out (to $3.2 \sigma$) a 1$\times$ solar composition atmosphere with 10 bars of surface pressure, and (to $3.1 \sigma$) the same atmosphere with with 1 bar of surface pressure.}
\label{fig:transmission}
\end{figure*}

We compare the model transmission spectra to the observed transmission spectrum (Figure~\ref{fig:transmission}). We disfavor a clear, solar composition atmospheres at 10 bars of surface pressure at $3.2 \sigma$ confidence, and 1 bar to $3.1 \sigma$. The observed transmission spectrum is consistent with tenuous solar composition atmospheres with 0.1 bar and 0.01 bar of surface pressure or higher mean molecular weight atmospheres dominated by heavier species such as H\tsub{2}O, CO\tsub{2}, or O\tsub{2}. It is also possible that \LTT b is devoid of an atmosphere.

The presence of an aerosol layer on \LTT b would also hinder our ability to detect its atmosphere. At low resolution an aerosol layer would act as the de facto bottom of the atmosphere, truncating any spectral features below it. This means that with our observed transmission spectrum we can rule out a high pressure (low altitude) aerosol layer at 1 bar or higher, but a low pressure (high-altitude) aerosol layer is degenerate with the cases of a tenuous (0.1 bar or less) atmosphere or no atmosphere. Probing the cases of an aerosol-rich atmosphere on \LTT b would require broader wavelength coverage or higher resolution spectral information.

We note that the transmission spectrum derived from Data Set 1 (blue bars in Figure~\ref{fig:indvspeclcs}) is discrepant from the others in several spectroscopic bands and appears to have a downward slope towards bluer wavelengths. Such slopes have been interpreted as evidence of unocculted bright spots, or faculae, on the surface of the star \citep[e.g.,][for the sub-Neptune GJ 1214b]{Rackham2017}. These effects are multiplicative with transit depth, and so unlikely to be observable in \LTT b's transmission spectrum at the precision we are able to achieve with Magellan/LDSS3C. We find a more plausible explanation in an overall increase in variability in the LTT 1445BC comparison star in Data Set 1 (see Section~\ref{subsec:Ha_variability}). As a test we remove Data Set 1 from the analysis, and find that we are able to rule out the 10 and 1 bar atmospheric cases to even higher confidence ($4.8 \sigma$ and $4.7 \sigma$, respectively).

\subsection{Flaring and H\texorpdfstring{$\alpha$}\\ variability in LTT 1445BC}\label{subsec:Ha_variability}

Analysis of two sectors of TESS data reveals that one or both of the BC components in the LTT 1445 system exhibits flare activity and \Ha\ in emission \citep{Winters2019,Winters2022}. Using X-ray time series observations by the Chandra X-ray Observatory, \citet{Brown2022} are able to resolve the A, B, and C, components of the LTT 1445 system and determine that C is the brightest X-ray source of the three stars, making it a likely candidate for the origin of the activity signal. We identify a flare in Data Set 1 originating from LTT 1445BC, as well as time-varying \Ha\ emission LTT 1445BC in all four data sets. We first investigate the \Ha\ emission from LTT 1445BC and then try to determine which component the flare and variability arise from. 

\subsubsection{LTT 1445BC as one source}

We construct 1) broadband lightcurves for LTT 1445BC that include the \Ha\ wavelengths (left out in the main analysis), and 2) narrow \Ha-band lightcurves to investigate the \Ha\ variability. To construct the LTT 1445BC lightcurves we use the primary star \LTT\ to remove telluric and instrumental variation from LTT 1445BC. \LTT\ is \Ha-quiet and does not exhibit any variability during the four observations. When removing systematics from these light curves we find that the GP regression is too flexible and removes some of the astrophysical variability we are trying to focus on, particularly in the \Ha-only light curves. We instead opt to use a simple linear decomposition using of all possible regression vectors provided in Table~\ref{tab:regressors} (similar to the analysis method described in \citet{Diamond-Lowe2018}). Though this process removes known correlated noise, some correlated noise from an unknown source remains in these light curves. We do not perform any outlier clipping of the light curve so as to preserve flare points. 

We remove the \LTT b transit from the \LTT\ light curve by fixing the transit parameters to those of the best-fit transit broadband transit (in the case of the broadband light curves) and the best-fit transit in the 640--660 nm spectroscopic band (for the \Ha\ light curves). We can then remove this exact transit shape from the \LTT\ light curve. After removing the best-fit transit and systematics model, we simply invert the lightcurve such that that BC component is divided by the A component, instead of the other way around. This way a flare from the BC component will appear as an increase in flux, instead of a decrease in flux.

\begin{figure}
\includegraphics[width=0.48\textwidth]{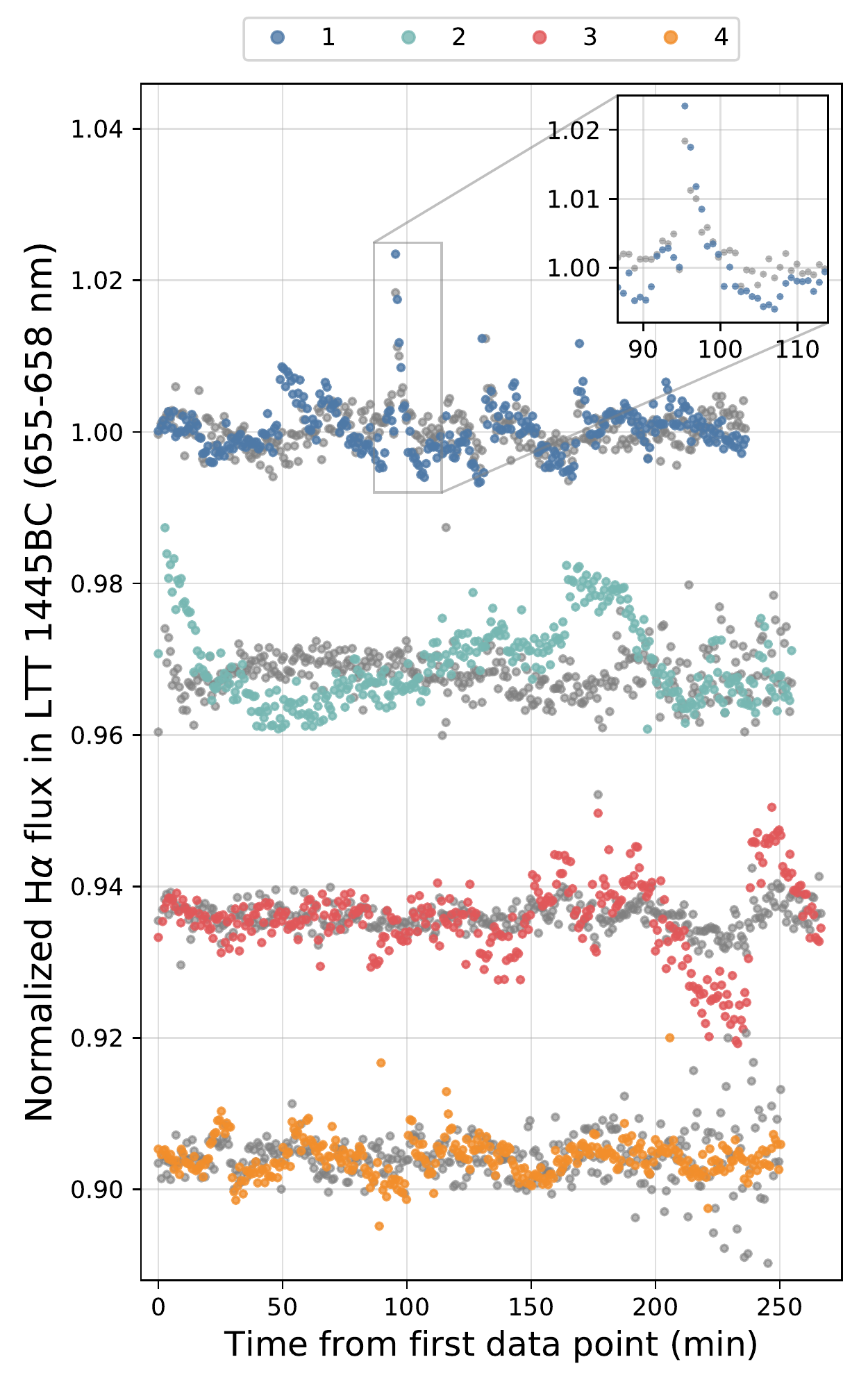}
\caption{Normalized \Ha\ (655--658 nm) light curves from LTT 1445BC in each of the four data sets analyzed in this work. Grey points in the background are the integrated broadband flux in each data set. The data are not flux calibrated so the light curves are presented as normalized flux. We multiply the broadband flux variability (grey points) by a factor of 10 so as to easily compare to the variability in the \Ha-only light curves. The inset shows a broadband flare identified both in Data Set 1.}
\label{fig:Halpha}
\end{figure}

In Figure~\ref{fig:Halpha} we present the variability in \Ha\ flux (655--658 nm) in Data Sets 1--4. For comparison, we plot the broadband light curves behind in grey. The data are not flux calibrated so we only present normalized \Ha\ and broadband light curves. The broadband data have inherently less scatter since we sum over all wavelengths, so we multiply the existing scatter by a factor of 10 in order to compare it more easily to the \Ha\ variability. It is clear that some variability in the \Ha\ flux is actually due to an overall trend in flux---for example, the first 20 minutes of Data Set 2 show a distinct slope in both the broadband and \Ha\ flux that was not removed by the linear regression. However, between minutes 150 and 200 of the same data set there is clear variability in the \Ha\ flux where the broadband flux remains stable. We also zoom in on the flare in Data Set 1, which occurs at all wavelengths. This is why, as stated in Section~\ref{sec:analysis}, we remove the data points associated with the flare in Data Set 1, and the \Ha\ wavelengths from all data sets.

We additionally search the combined BC spectra for emission and variability in other known bands related to activity indicators, such as K \textsc{i} (7664.90\AA, 7698.96\AA), Na \textsc{i} (8183.26\AA, 8194.82\AA), and Ca \textsc{ii} (8498.02\AA, 8542.09\AA, 8662.14\AA), which are detected in the more active M dwarf Proxima Centauri \citep{Robertson2016}. We do not see evidence of these lines, either because LTT 1445BC lacks these indicators, or because our data are taken at relatively low resolution ($R\sim~1000$). In either case, we see no evidence of additional variability from the BC component imprinting on the spectroscopic light curves and resulting transit depths.

\subsubsection{Separating LTT 1445B and C in H$\alpha$}

\begin{figure*}
\includegraphics[width=0.98\textwidth]{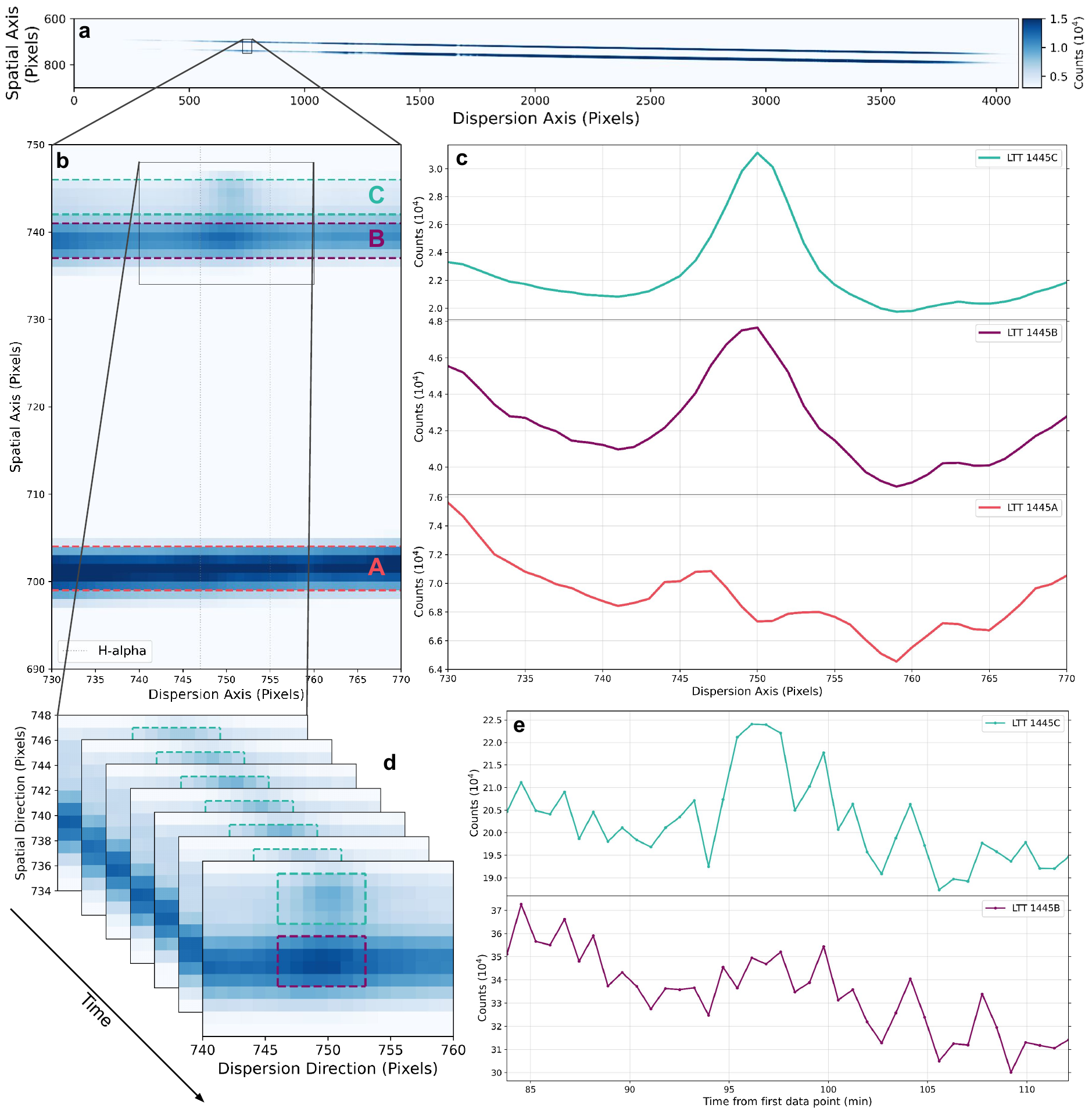}
\caption{Steps to separate the \Ha\ spectra in time series of LTT 1445B and C from Data Set 1. Panel a: Stacked spectrum of raw images from Data Set 1. Panel b: Zoomed in to the spectral region around the \Ha\ line. \Ha\ emission is visible by eye in stars B and C, but not in A. Panel c: Pixels are summed in the spatial direction, as indicated by the dashed lines in Panel b, and plotted along the dispersion axis. \Ha\ is present in emission for both LTT 1445 B and C. Panel d: We sum the counts in a box of pixels around the \Ha\ lines in both B and C as indicated by the dashed lines. We do this for each exposure across the time span where we see the flare in Data Set 1. Panel e: Counts as a function of time for B and C. It is clear that the flare in Data Set 1 occurring around the 95\tsup{th} minute after the start of the time series originates from LTT 1445C.}
\label{fig:Halphaseparate}
\end{figure*}

The spectra of LTT 1445B and C cannot be fully separated in the Magellan II/LDSS3C data, especially in the redder part of the spectrum ($>700$ nm in Figure~\ref{fig:spectrabinned}) where the PSF of the two stars in the spatial direction becomes completely blended. However, the \Ha\ line at the blue end of the spectra where M dwarfs emit few photons, can be seen by eye in the raw data (Figure~\ref{fig:Halphaseparate}). From this raw data we sum up the rows in the spatial direction associated with each star and plot the spectrum. \Ha\ can be seen in emission for both LTT 1445B and C, and is absent from \LTT, as expected. 

We then determine which component, B or C, is responsible for the flare in Data Set 1. We take the raw exposures around the time of the flare and sum up the flux in a box of pixels centered on the \Ha\ line for B and C, and can determine that LTT 1445C is the origin of the flare in Data Set 1 (Figure~\ref{fig:Halphaseparate}). Looking at the whole time series, we additionally find that C is responsible for the bulk of the \Ha\ variability in Data Set 1, and that Data Set 1 exhibits overall more variability than Data Sets 2, 3, and 4.

Time-varying \Ha\ emission is an well-measured phenomenon in M dwarfs \citep[e.g.,][]{Lee2010}. High levels of \Ha\ activity are correlated with rapid stellar rotation periods \citep{Newton2016a,Newton2018,Medina2020}, though \Ha\ variability has not been shown to correlate with rotational phase \citep{Medina2022}. Instead, \citet{Medina2022} posit that low-energy flares are responsible for the observed \Ha\ variation. The variation seen in the Magellan II/LDSS3C \Ha\ light curves occurs on timescales of minutes to hours, and is ultimately undetectable in the broadband light curves (which cover approximately the same spectral range as TESS). It is only when we zoom in to the \Ha\ band that this rich variability revealed, suggesting that short time scale \Ha\ variability is hidden in the broadband photometric TESS data, especially since the more active LTT~1445C is fainter than B at optical wavelengths. That LTT 1445C is the more active component is in agreement with rapid rotation rate estimates \citep{Winters2019}, and with X-ray data showing that LTT 1445C is the brightest X-ray source in the system \citep{Brown2022}.

Based on two sectors of TESS data (Sectors 4 and 31), as well as a year of ground-based photometric monitoring of the LTT 1445 system by MEarth \citep{Nutzman2008,Irwin2015}, \citet{Winters2022} estimate a 1.4 day rotation period for LTT 1445C and perhaps a 6.7 day rotation period for LTT 1445B. While our narrow-band \Ha\ observations demonstrate clear variability, these observations span 4.2 hr, or about 12.5\% of the dominant 1.4 day rotation period seen in the TESS data. The four Magellan/LDSS3C data sets presented here are taken from one month to one year apart (Table~\ref{tab:obs}), so these data do not represent consistent time coverage comparable to the TESS data. The relatively short time coverage of the Magellan II/LDSS3C data is too sparse to further constrain the larger rotational variation of LTT 1445B or C.

\section{Conclusion}\label{sec:conclusion}

We observed four transits of the nearby exoplanet \LTT b using the ground-based multi-object spectrograph LDSS3C mounted on the Magellan II (Clay) telescope at the Las Campanas Observatory in Chile. A particular advantage of observing the LTT 1445 system is that the close binary pair LTT 1445BC can be used as a comparison star to \LTT\ when removing telluric variability during observations. \LTT\ is well separated from the LTT 1445BC binary pair, but we are unable to fully resolve the spectra of the B and C components in our data. However, we do determine that LTT 1445C is responsible for the bulk of the \Ha\ variation seen in the LTT 1445BC light curves, as well as for a flare detected in Data Set 1. When constructing broadband and spectroscopic light curves for analysis we exclude the flare time points from the Data Set 1 and all spectral data in the \Ha\ band (655--680 nm).

With the data presented in this work we rule out a clear, low mean molecular weight atmosphere at 10 bars of surface pressure ($3.2 \sigma$) and 1 bar of surface pressure ($3.1 \sigma$). This result joins previous work on GJ 1132b \citep{Diamond-Lowe2018,Mugnai2021,Libby-Roberts2022}, TRAPPIST-1a--f \citep{deWit2016,deWit2018}, LHS 3844 \citep{Diamond-Lowe2020b}, L 98-59b--d \citep{Damiano2022,Zhou2022a,Zhou2022b,Barclay2023}, and LHS 475b \citep{Lustig-Yaeger2023} to strongly suggest that highly irradiated terrestrial exoplanets orbiting M dwarfs are not capable of retaining low mean molecular weight atmospheres, if they are able to accrete them in the first place. It is possible, however, that \LTT b possesses a high mean molecular weight or cloudy/hazy atmosphere, which would fall below our detection limits, or no atmosphere at all. To determine whether or not a compact, high mean molecular weight atmosphere is present around this world, JWST Cycle 1 GO Program 2708 (PI Z.\ Berta-Thompson) will observe secondary eclipses of this world at mid-infrared wavelengths and search for signs of energy advection that would imply the presence of an atmosphere.

\vspace{.5cm}
%\begin{acknowledgments}
This work is made possible by the dedicated astronomers, telescope operators, instrument specialists, engineers, administrators, and support staff at Las Campanas Observatory. We are grateful to Yuri Beletsky for executing the observations used in this work. We thank Dave Osip and Andreas Seifahrt for extensive conversations about the capability of Magellan II and LDSS3C to collect these data. We thank Jennifer Winters for helpful communications regarding stellar and planetary parameters of \LTT\ and \LTT b, Amber Medina for fruitful discussions about \Ha\ variability in M dwarfs, and Jonathan Irwin for contributions to the original observing proposal. We thank Matej Malik for sharing the custom opacity tables used in this work. We thank the Exoplanet Group at DTU Space for insightful conversations during the preparation of this manuscript. We thank the anonymous referee for their comments.

This research has made use of the Exoplanet Follow-up Observation Program website and the NASA Exoplanet Archive, which are operated by the California Institute of Technology, under contract with the National Aeronautics and Space Administration under the Exoplanet Exploration Program. The data analysis described in this work was performed with the help of the high performance cluster at the DTU Computing Center \citep{DCC_HPC}. This material is based upon work supported by the National Aeronautics and Space Administration under grant No.\ 80NSSC18K0476 issued through the XRP Program. This project is partly funded by Villum Fonden. 

%\end{acknowledgments}

%\vspace{5mm}
\facility{Magellan:Clay (LDSS3C multi-object spectrograph)}

\software{\texttt{astropy} \citep{AstropyCollaboration2013,AstropyCollaboration2018}, 
         \texttt{batman} \citep{Kreidberg2015}, 
         \texttt{decorrasaurus} \citep{Diamond-Lowe2020a}, 
         \texttt{dynesty} \citep{Speagle2020},
         \texttt{Exo-Transmit} \citep{Kempton2017},
         \texttt{FastChem} \citep{Stock2018},
         \texttt{george} \citep{Ambikasaran2015},
         \texttt{HELIOS-K} \citep{Grimm2021},
         \texttt{LDTk} \citep{Parviainen2015}, 
         \texttt{mosasaurus} (\href{http://www.github.com/zkbt/mosasaurus}{github.com/zkbt/mosasaurus})}

\bibliography{MasterBibliography}

\end{document}